\documentclass[12pt]{article}

\usepackage[letterpaper, margin=1in]{geometry}

\usepackage[colorlinks]{hyperref}
\usepackage{makeidx}
\usepackage{framed}
\usepackage{fancyvrb}
\usepackage{color}
\usepackage{hyperref}
\hypersetup{colorlinks = true, allcolors = blue}
\usepackage{float}
\usepackage{flafter}
\usepackage{tabularx}
\usepackage{booktabs}
\usepackage{longtable}
\usepackage{multirow}
\usepackage{graphicx}
\usepackage{afterpage}

\usepackage{datetime}

\yyyymmdddate

  \usepackage{setspace}
  \AtBeginEnvironment{tabular}{\singlespacing}
  \AtBeginEnvironment{table}{\singlespacing}
  \AtBeginEnvironment{longtable}{\singlespacing}
  \AtBeginEnvironment{verbatim}{\singlespacing}
  \AtBeginEnvironment{Shaded}{\singlespacing}

\NewDocumentCommand\citeproctext{}{}

\makeatletter
 \let\@cite@ofmt\@firstofone
 \def\@biblabel#1{}
 \def\@cite#1#2{{#1\if@tempswa , #2\fi}}
\makeatother
\newlength{\cslhangindent}
\newlength{\csllabelwidth}
\newenvironment{CSLReferences}[2] 
 {\begin{list}{}{%
  \setlength{\itemindent}{0pt}
  \setlength{\leftmargin}{0pt}
  \setlength{\parsep}{0pt}
  \ifodd #1
   \setlength{\leftmargin}{\cslhangindent}
   \setlength{\itemindent}{-1\cslhangindent}
  \fi
  \setlength{\itemsep}{#2\baselineskip}}}
 {\end{list}}
\usepackage{calc}

  \usepackage{amsmath}
  \usepackage{amssymb}

\usepackage{xcolor}
\definecolor{dark}{HTML}{2c2e35}
\color{dark}

\definecolor{myblue}{HTML}{1e3765}
\hypersetup{colorlinks,linkcolor=myblue,urlcolor=myblue}

\makeindex

\providecommand{\tightlist}{\setlength{\itemsep}{0pt}\setlength{\parskip}{0pt}}

  \usepackage{color}
  \usepackage{fancyvrb}

  \DefineVerbatimEnvironment{Highlighting}{Verbatim}{commandchars=\\\{\}}
  \usepackage{framed}
  \definecolor{shadecolor}{RGB}{241,243,245}
  \newenvironment{Shaded}{\begin{snugshade}}{\end{snugshade}}

  \newcommand{\AttributeTok}[1]{\textcolor[rgb]{0.40,0.45,0.13}{#1}}
  
  \newcommand{\BuiltInTok}[1]{\textcolor[rgb]{0.00,0.23,0.31}{#1}}
  
  \newcommand{\CommentTok}[1]{\textcolor[rgb]{0.37,0.37,0.37}{#1}}

  \newcommand{\ControlFlowTok}[1]{\textcolor[rgb]{0.00,0.23,0.31}{\textbf{#1}}}
  \newcommand{\DataTypeTok}[1]{\textcolor[rgb]{0.68,0.00,0.00}{#1}}
  \newcommand{\DecValTok}[1]{\textcolor[rgb]{0.68,0.00,0.00}{#1}}

  \newcommand{\FunctionTok}[1]{\textcolor[rgb]{0.28,0.35,0.67}{#1}}
  \newcommand{\ImportTok}[1]{\textcolor[rgb]{0.00,0.46,0.62}{#1}}
  
  \newcommand{\KeywordTok}[1]{\textcolor[rgb]{0.00,0.23,0.31}{\textbf{#1}}}
  \newcommand{\NormalTok}[1]{\textcolor[rgb]{0.00,0.23,0.31}{#1}}
  \newcommand{\OperatorTok}[1]{\textcolor[rgb]{0.37,0.37,0.37}{#1}}
  \newcommand{\OtherTok}[1]{\textcolor[rgb]{0.00,0.23,0.31}{#1}}
  \newcommand{\PreprocessorTok}[1]{\textcolor[rgb]{0.68,0.00,0.00}{#1}}
  
  \newcommand{\SpecialCharTok}[1]{\textcolor[rgb]{0.37,0.37,0.37}{#1}}
  
  \newcommand{\StringTok}[1]{\textcolor[rgb]{0.13,0.47,0.30}{#1}}
  \newcommand{\VariableTok}[1]{\textcolor[rgb]{0.07,0.07,0.07}{#1}}

\author{
  Mauricio Vargas Sepúlveda (ORCID 0000-0003-1017-7574)\\Department of
Political Science, University of Toronto\\Munk School of Global Affairs
and Public Policy, University of Toronto\\
  \smallskip\\
  Corresponding author: m.sepulveda@mail.utoronto.ca
}
\title{Welding R and C++: A Tale of Two Programming Languages}
\date{Last updated: \today\ \currenttime}

\begin{document}

\maketitle

\thispagestyle{empty}
\tableofcontents
\setcounter{page}{0}
\clearpage

\afterpage{\setlength\parskip{10pt}}

\section{Abstract}\label{abstract}

This article compares \texttt{cpp11armadillo} and \texttt{cpp11eigen},
new R packages that integrate the powerful Armadillo and Eigen C++
libraries for linear algebra into the R programming environment. This
article provides a detailed comparison between Armadillo and Eigen speed
and syntax. The goal of these packages is to simplify a part of the
process of solving bottlenecks by using C++ within R, these offer
additional ease of integration for users who require high-performance
linear algebra operations in their R workflows. This document aims to
discuss the tradeoff between computational efficiency and accessibility.

\section{Introduction}\label{introduction}

R is widely used by non-programmers (Wickham et al. 2019), and this
article aims to introduce benchmarks in a non-technical yet formal
manner for social scientists. Our goal is to provide a fair comparison
between Eigen and Armadillo, being both highly efficient linear algebra
libraries written in C++. We do it by using
\href{https://pacha.dev/cpp11armadillo}{\texttt{cpp11armadillo}} and
\href{https://pacha.dev/cpp11eigen}{\texttt{cpp11eigen}}.

\href{https://arma.sourceforge.net/}{Armadillo} is a C++ library
designed for linear algebra, emphasizing a balance between performance
and ease of use. C++ is highly efficient for computationally intensive
tasks but lacks built-in data structures and functions for linear
algebra operations. Armadillo fills this gap by providing an intuitive
syntax similar to MATLAB (Sanderson and Curtin 2016).

\href{https://eigen.tuxfamily.org/index.php?title=Main_Page}{Eigen}
emphasizes flexibility and speed, while
\href{http://arma.sourceforge.net/}{Armadillo} focuses on a balance
between speed and easy of use.

\href{https://cran.r-project.org/package=RcppArmadillo}{\texttt{RcppArmadillo}},
introduced in 2010, integrates R and Armadillo (Sanderson and Curtin
2016; Eddelbuettel and Sanderson 2014).
\href{https://cran.r-project.org/package=RcppEigen}{\texttt{RcppEigen}},
introduced in 2011, integrates Eigen with R through the \texttt{Rcpp}
package introduced in 2008, enabling the use of C++ for
performance-critical parts of R code. At the time of writing this
document, 732 CRAN packages depend on \texttt{RcppArmadillo} and 238 on
\texttt{RcppEigen} (Lee 2024), and therefore these are highly successful
packages considering that the median number of reverse dependencies for
CRAN packages is five and the distribution of dependencies situates
these packages as top one percent.

\texttt{cpp11armadillo} and \texttt{cpp11eigen} are independent project
that aim to simplify the integration of R and C++ by using
\texttt{cpp11}, an R package introduced in 2020 that eases calling C++
functions from R, it is currently used as a dependency by 75 CRAN
packages, and it is in the top one percent of CRAN packages (R Core Team
2024).

A distinctive characteristic of \texttt{cpp11armadillo} and
\texttt{cpp11eigen} is the vendoring capability, meaning that it allows
to copy its code into a project, making it a one-time dependency with a
fixed and stable code until it is manually updated. This feature is
useful in restricted environments such as servers and clusters where
sometimes there are restrictions to software installation or the
internet connections are limited for security reasons (Wickham et al.
2019; Vaughan, Hester, and François 2023).

\texttt{cpp11armadillo} and \texttt{cpp11eigen} are useful in cases
where vectorization (e.g., applying an operation to a vector or matrix
as a whole instead of looping over each element) is not possible or
challenging. A detailed discussion and examples about why and when (and
when not) rewriting R code in C++ is useful can be found in Burns
(2011). We followed four design principles when developing these two
packages as in: column oriented, package oriented, header-only, and
vendoring capable. The details of the \texttt{cpp11armadillo}
implementation, which is similar to \texttt{cpp11eigen}, can be found in
Vargas Sepúlveda and Schneider Malamud (2024).

\section{Syntax}\label{syntax}

One possibility is to start by creating minimal R packages with the
provided templates. These templates provide a general case for a package
that includes the necessary files to create a package that uses
Armadillo or Eigen, including a generic \texttt{Makevars} file that can
be adapted to link to specific numerical libraries such as Intel MKL or
OpenBLAS.

\begin{Shaded}
\begin{Highlighting}[]
\NormalTok{remotes}\SpecialCharTok{::}\FunctionTok{install\_github}\NormalTok{(}\StringTok{"pachadotdev/cpp11armadillo"}\NormalTok{)}
\NormalTok{remotes}\SpecialCharTok{::}\FunctionTok{install\_github}\NormalTok{(}\StringTok{"pachadotdev/cpp11eigen"}\NormalTok{)}

\NormalTok{cpp11eigen}\SpecialCharTok{::}\FunctionTok{create\_package}\NormalTok{(}\StringTok{"armadillobenchmark"}\NormalTok{)}
\NormalTok{cpp11eigen}\SpecialCharTok{::}\FunctionTok{create\_package}\NormalTok{(}\StringTok{"eigenbenchmark"}\NormalTok{)}
\end{Highlighting}
\end{Shaded}

Comparing numerical libraries requires to write equivalent codes. For
instance, in R we use \texttt{apply()} while its C++ equivalent is a
\texttt{for} loop, and this allows a fair comparison between the two
libraries. However, R has heavily optimized functions that also verify
the input data, such as \texttt{lm()} and \texttt{glm()}, which do not
have a direct equivalent in Armadillo or Eigen, and for a fair
comparison the options are to write a simplified function for the linear
model in R or to write a more complex function in C++.

The ATT benchmark, is a set of functions that can be rewritten using
Armadillo and Eigen with relative ease, and test has the advantage of
being well-known and widely used in the R community.

The first test in the ATT benchmark is the creation, transposition and
deformation of an \(N \times N\) matrix (\(2,500 \times 2,500\) in the
original test). The R and Armadillo codes for this operation are:

\begin{Shaded}
\begin{Highlighting}[]
\CommentTok{\# R}

\NormalTok{matrix\_calculation\_01\_r }\OtherTok{\textless{}{-}} \ControlFlowTok{function}\NormalTok{(n) \{}
\NormalTok{  a }\OtherTok{\textless{}{-}} \FunctionTok{matrix}\NormalTok{(}\FunctionTok{rnorm}\NormalTok{(n }\SpecialCharTok{*}\NormalTok{ n) }\SpecialCharTok{/} \DecValTok{10}\NormalTok{, }\AttributeTok{ncol =}\NormalTok{ n, }\AttributeTok{nrow =}\NormalTok{ n)}
\NormalTok{  b }\OtherTok{\textless{}{-}} \FunctionTok{t}\NormalTok{(a)}
  \FunctionTok{dim}\NormalTok{(b) }\OtherTok{\textless{}{-}} \FunctionTok{c}\NormalTok{(n }\SpecialCharTok{/} \DecValTok{2}\NormalTok{, n }\SpecialCharTok{*} \DecValTok{2}\NormalTok{)}
\NormalTok{  a }\OtherTok{\textless{}{-}} \FunctionTok{t}\NormalTok{(b)}
  \FunctionTok{return}\NormalTok{(}\DecValTok{0}\NormalTok{L)}
\NormalTok{\}}
\end{Highlighting}
\end{Shaded}

\begin{Shaded}
\begin{Highlighting}[]
\CommentTok{// C++}

\PreprocessorTok{\#include }\ImportTok{\textless{}cpp11.hpp\textgreater{}}
\PreprocessorTok{\#include }\ImportTok{\textless{}cpp11armadillo.hpp\textgreater{}}

\KeywordTok{using} \KeywordTok{namespace}\NormalTok{ arma}\OperatorTok{;}
\KeywordTok{using} \KeywordTok{namespace}\NormalTok{ cpp11}\OperatorTok{;}

\OperatorTok{[[}\AttributeTok{cpp11}\OperatorTok{::}\AttributeTok{register}\OperatorTok{]]} \DataTypeTok{int} \VariableTok{matrix\_calculation\_01\_arma\_}\OperatorTok{(}\AttributeTok{const} \DataTypeTok{int}\OperatorTok{\&}\NormalTok{ n}\OperatorTok{)} \OperatorTok{\{}
\NormalTok{  mat a }\OperatorTok{=}\NormalTok{ randn}\OperatorTok{\textless{}}\NormalTok{mat}\OperatorTok{\textgreater{}(}\NormalTok{n}\OperatorTok{,}\NormalTok{n}\OperatorTok{)} \OperatorTok{/} \DecValTok{10}\OperatorTok{;}
\NormalTok{  mat b }\OperatorTok{=}\NormalTok{ a}\OperatorTok{.}\NormalTok{t}\OperatorTok{();}
\NormalTok{  b}\OperatorTok{.}\NormalTok{reshape}\OperatorTok{(}\NormalTok{n}\OperatorTok{/}\DecValTok{2}\OperatorTok{,}\NormalTok{ n}\OperatorTok{*}\DecValTok{2}\OperatorTok{);}
\NormalTok{  a }\OperatorTok{=}\NormalTok{ b}\OperatorTok{.}\NormalTok{t}\OperatorTok{();}
  \ControlFlowTok{return} \DecValTok{0}\OperatorTok{;}
\OperatorTok{\}}
\end{Highlighting}
\end{Shaded}

The Eigen code requires to create a function to draw random numbers from
a normal distribution because it only provides a built-in function for
the uniform distribution:

\begin{Shaded}
\begin{Highlighting}[]
\PreprocessorTok{\#include }\ImportTok{\textless{}cpp11.hpp\textgreater{}}
\PreprocessorTok{\#include }\ImportTok{\textless{}cpp11eigen.hpp\textgreater{}}
\PreprocessorTok{\#include }\ImportTok{\textless{}random\textgreater{}}

\KeywordTok{using} \KeywordTok{namespace}\NormalTok{ Eigen}\OperatorTok{;}
\KeywordTok{using} \KeywordTok{namespace}\NormalTok{ cpp11}\OperatorTok{;}

\BuiltInTok{std::}\NormalTok{mt19937}\OperatorTok{\&}\NormalTok{ random\_normal}\OperatorTok{()} \OperatorTok{\{}
  \AttributeTok{static} \BuiltInTok{std::}\NormalTok{random\_device rd}\OperatorTok{;}
  \AttributeTok{static} \BuiltInTok{std::}\NormalTok{mt19937 gen}\OperatorTok{(}\NormalTok{rd}\OperatorTok{());}
  \ControlFlowTok{return}\NormalTok{ gen}\OperatorTok{;}
\OperatorTok{\}}

\OperatorTok{[[}\AttributeTok{cpp11}\OperatorTok{::}\AttributeTok{register}\OperatorTok{]]} \DataTypeTok{int} \VariableTok{matrix\_calculation\_01\_eigen\_}\OperatorTok{(}\AttributeTok{const} \DataTypeTok{int}\OperatorTok{\&}\NormalTok{ n}\OperatorTok{)} \OperatorTok{\{}
  \BuiltInTok{std::}\NormalTok{normal\_distribution}\OperatorTok{\textless{}}\DataTypeTok{double}\OperatorTok{\textgreater{}}\NormalTok{ d}\OperatorTok{(}\DecValTok{0}\OperatorTok{,} \DecValTok{1}\OperatorTok{);}
  
\NormalTok{  MatrixXd a }\OperatorTok{=}\NormalTok{ MatrixXd}\OperatorTok{::}\NormalTok{NullaryExpr}\OperatorTok{(}\NormalTok{n}\OperatorTok{,}\NormalTok{ n}\OperatorTok{,} \OperatorTok{[\&]()} \OperatorTok{\{}
    \ControlFlowTok{return}\NormalTok{ d}\OperatorTok{(}\NormalTok{random\_normal}\OperatorTok{());}
  \OperatorTok{\})} \OperatorTok{/} \DecValTok{10}\OperatorTok{;}

  \CommentTok{// for the uniform distribution this is just}
  \CommentTok{// MatrixXd a = MatrixXd::Random(n, n) / 10;}

\NormalTok{  MatrixXd b }\OperatorTok{=}\NormalTok{ a}\OperatorTok{.}\NormalTok{transpose}\OperatorTok{();}
\NormalTok{  b}\OperatorTok{.}\NormalTok{resize}\OperatorTok{(}\NormalTok{n }\OperatorTok{/} \DecValTok{2}\OperatorTok{,}\NormalTok{ n }\OperatorTok{*} \DecValTok{2}\OperatorTok{);}
  \ControlFlowTok{return} \DecValTok{0}\OperatorTok{;}
\OperatorTok{\}}
\end{Highlighting}
\end{Shaded}

\section{Benchmarks without data
transfer}\label{benchmarks-without-data-transfer}

The functions in the previous section to do not move data between R and
C++, this is intentional to focus on the performance of the linear
algebra libraries and not adding overhead from data transfer in the
benchmarks. Each function creates a matrix and conducts equivalent
operations on it. The returned value is zero in R and C++ in case that
the functions run without errors.

We decided to run the benchmarks on a local machine and on a cluster to
provide a comparison between the two environments and test how the
benchmarks change when the input data is increased.

\subsection{Local benchmarks}\label{local-benchmarks}

The local benchmarks were conducted on a ThinkPad X1 Carbon Gen 9 with
the following specifications:

\begin{itemize}
\tightlist
\item
  Processor: Intel Core i7-1185G7 with eight cores
\item
  Memory: 16 GB LPDDR4Xx-4266
\item
  Operating System: Pop!\_OS 22.04 based on Ubuntu 22.04
\item
  R Version: 4.4.1
\item
  BLAS Library: OpenBLAS 0.3.20
\end{itemize}

The median times for the adapted and comparable implementations of the
ATT benchmarks are as follows:

\begin{longtable}[]{@{}
  >{\raggedright\arraybackslash}p{(\columnwidth - 4\tabcolsep) * \real{0.7717}}
  >{\raggedleft\arraybackslash}p{(\columnwidth - 4\tabcolsep) * \real{0.1739}}
  >{\raggedleft\arraybackslash}p{(\columnwidth - 4\tabcolsep) * \real{0.0543}}@{}}
\caption{Matrix calculation}\tabularnewline
\toprule\noalign{}
\begin{minipage}[b]{\linewidth}\raggedright
Operation
\end{minipage} & \begin{minipage}[b]{\linewidth}\raggedleft
Median time (s)
\end{minipage} & \begin{minipage}[b]{\linewidth}\raggedleft
Rank
\end{minipage} \\
\midrule\noalign{}
\endfirsthead
\toprule\noalign{}
\begin{minipage}[b]{\linewidth}\raggedright
Operation
\end{minipage} & \begin{minipage}[b]{\linewidth}\raggedleft
Median time (s)
\end{minipage} & \begin{minipage}[b]{\linewidth}\raggedleft
Rank
\end{minipage} \\
\midrule\noalign{}
\endhead
\bottomrule\noalign{}
\endlastfoot
\(2,400 \times 2,400\) \(\text{matrix}^{1,000}\) - Armadillo & \(0.188\)
& 1 \\
\(2,400 \times 2,400\) \(\text{matrix}^{1,000}\) - Eigen & \(0.301\) &
2 \\
\(2,400 \times 2,400\) \(\text{matrix}^{1,000}\) - R & \(0.325\) & 3 \\
\(2,800 \times 2,800\) cross-product matrix - Armadillo & \(0.398\) &
1 \\
\(2,800 \times 2,800\) cross-product matrix - R & \(0.444\) & 2 \\
\(2,800 \times 2,800\) cross-product matrix - Eigen & \(1.151\) & 3 \\
Creation and modification of a \(2,500 \times 2,500\) matrix - Armadillo
& \(0.204\) & 1 \\
Creation and modification of a \(2,500 \times 2,500\) matrix - Eigen &
\(0.232\) & 2 \\
Creation and modification of a \(2,500 \times 2,500\) matrix - R &
\(0.294\) & 3 \\
Linear regression over a \(3,000 \times 3,000\) matrix - Armadillo &
\(0.459\) & 1 \\
Linear regression over a \(3,000 \times 3,000\) matrix - R & \(5.303\) &
2 \\
Linear regression over a \(3,000 \times 3,000\) matrix - Eigen &
\(8.809\) & 3 \\
Sorting of 7,000,000 values - Armadillo & \(0.663\) & 1 \\
Sorting of 7,000,000 values - Eigen & \(0.691\) & 2 \\
Sorting of 7,000,000 values - R & \(0.759\) & 3 \\
\end{longtable}

\begin{longtable}[]{@{}
  >{\raggedright\arraybackslash}p{(\columnwidth - 4\tabcolsep) * \real{0.7640}}
  >{\raggedleft\arraybackslash}p{(\columnwidth - 4\tabcolsep) * \real{0.1798}}
  >{\raggedleft\arraybackslash}p{(\columnwidth - 4\tabcolsep) * \real{0.0562}}@{}}
\caption{Matrix functions}\tabularnewline
\toprule\noalign{}
\begin{minipage}[b]{\linewidth}\raggedright
Operation
\end{minipage} & \begin{minipage}[b]{\linewidth}\raggedleft
Median time (s)
\end{minipage} & \begin{minipage}[b]{\linewidth}\raggedleft
Rank
\end{minipage} \\
\midrule\noalign{}
\endfirsthead
\toprule\noalign{}
\begin{minipage}[b]{\linewidth}\raggedright
Operation
\end{minipage} & \begin{minipage}[b]{\linewidth}\raggedleft
Median time (s)
\end{minipage} & \begin{minipage}[b]{\linewidth}\raggedleft
Rank
\end{minipage} \\
\midrule\noalign{}
\endhead
\bottomrule\noalign{}
\endlastfoot
Cholesky decomposition of a \(3,000 \times 3,000\) matrix - Armadillo &
\(0.608\) & 1 \\
Cholesky decomposition of a \(3,000 \times 3,000\) matrix - R &
\(0.709\) & 2 \\
Cholesky decomposition of a \(3,000 \times 3,000\) matrix - Eigen &
\(2.902\) & 3 \\
Determinant of a \(2,500 \times 2,500\) matrix - Armadillo & \(0.293\) &
1 \\
Determinant of a \(2,500 \times 2,500\) matrix - R & \(0.303\) & 2 \\
Determinant of a \(2,500 \times 2,500\) matrix - Eigen & \(0.562\) &
3 \\
Eigenvalues of a \(640 \times 640\) matrix - Armadillo & \(0.367\) &
1 \\
Eigenvalues of a \(640 \times 640\) matrix - R & \(0.369\) & 2 \\
Eigenvalues of a \(640 \times 640\) matrix - Eigen & \(1.629\) & 3 \\
Fast Fourier Transform over 2,400,000 values - Eigen & \(0.14\) & 1 \\
Fast Fourier Transform over 2,400,000 values - R & \(0.23\) & 2 \\
Fast Fourier Transform over 2,400,000 values - Armadillo & \(0.294\) &
3 \\
Inverse of a \(1,600 \times 1,600\) matrix - Armadillo & \(0.312\) &
1 \\
Inverse of a \(1,600 \times 1,600\) matrix - R & \(0.324\) & 2 \\
Inverse of a \(1,600 \times 1,600\) matrix - Eigen & \(0.758\) & 3 \\
\end{longtable}

\begin{longtable}[]{@{}
  >{\raggedright\arraybackslash}p{(\columnwidth - 4\tabcolsep) * \real{0.7045}}
  >{\raggedleft\arraybackslash}p{(\columnwidth - 4\tabcolsep) * \real{0.2386}}
  >{\raggedleft\arraybackslash}p{(\columnwidth - 4\tabcolsep) * \real{0.0568}}@{}}
\caption{Programmation}\tabularnewline
\toprule\noalign{}
\begin{minipage}[b]{\linewidth}\raggedright
Operation
\end{minipage} & \begin{minipage}[b]{\linewidth}\raggedleft
Median time (s)
\end{minipage} & \begin{minipage}[b]{\linewidth}\raggedleft
Rank
\end{minipage} \\
\midrule\noalign{}
\endfirsthead
\toprule\noalign{}
\begin{minipage}[b]{\linewidth}\raggedright
Operation
\end{minipage} & \begin{minipage}[b]{\linewidth}\raggedleft
Median time (s)
\end{minipage} & \begin{minipage}[b]{\linewidth}\raggedleft
Rank
\end{minipage} \\
\midrule\noalign{}
\endhead
\bottomrule\noalign{}
\endlastfoot
3,500,000 Fibonacci numbers calculation - Eigen & \(1.4 \times 10^{-1}\)
& 1 \\
3,500,000 Fibonacci numbers calculation - Armadillo &
\(1.7 \times 10^{-1}\) & 2 \\
3,500,000 Fibonacci numbers calculation - R & \(1.7 \times 10^{-1}\) &
3 \\
Creation of a \(3,000 \times 3,000\) Hilbert matrix - Eigen &
\(4.6 \times 10^{-6}\) & 1 \\
Creation of a \(3,000 \times 3,000\) Hilbert matrix - Armadillo &
\(5.9 \times 10^{-2}\) & 2 \\
Creation of a \(3,000 \times 3,000\) Hilbert matrix - R &
\(1.5 \times 10^{-1}\) & 3 \\
Creation of a \(500 \times 500\) Toeplitz matrix - Eigen &
\(7.9 \times 10^{-7}\) & 1 \\
Creation of a \(500 \times 500\) Toeplitz matrix - Armadillo &
\(4 \times 10^{-4}\) & 2 \\
Creation of a \(500 \times 500\) Toeplitz matrix - R &
\(2.6 \times 10^{-3}\) & 3 \\
Escoufier's method on a \(45 \times 45\) matrix - Armadillo &
\(2.4 \times 10^{-2}\) & 1 \\
Escoufier's method on a \(45 \times 45\) matrix - Eigen &
\(3.2 \times 10^{-2}\) & 2 \\
Escoufier's method on a \(45 \times 45\) matrix - R &
\(1.4 \times 10^{-1}\) & 3 \\
Grand common divisors of 400,000 pairs - Eigen & \(2.1 \times 10^{-2}\)
& 1 \\
Grand common divisors of 400,000 pairs - Armadillo &
\(2.3 \times 10^{-2}\) & 2 \\
Grand common divisors of 400,000 pairs - R & \(1.884\) & 3 \\
\end{longtable}

The results reveal that Armadillo leads in most of the benchmarks, but
Eigen is particularly faster in some tests such as the Fast Fourier
Transform. R is the second or third in all benchmarks, but it is
important to note that R comes with an additional advantage in terms of
simplified syntax and the ability to run the code without compiling it.

These tests are not exhaustive, and we must be cautious when
interpreting the results. The ATT benchmark is a good starting point,
but it does not cover mundane tasks such as data manipulation, and it is
important to consider the tradeoff between computational efficiency and
ease of use.

\subsection{Cluster benchmarks}\label{cluster-benchmarks}

The cluster benchmarks were conducted on one cluster node of the
\href{https://docs.alliancecan.ca/wiki/Niagara}{Niagara supercomputer}
maintained by the Digital Research Alliance of Canada, which has the
following specifications:

\begin{itemize}
\tightlist
\item
  Processor: 2 sockets with 20 Intel Skylake cores (2.4GHz, AVX512), for
  a total of 40 cores per node
\item
  Memory: 202 GB
\item
  Operating System: CentOS 7
\item
  R Version: 4.2.2
\item
  BLAS Library: Intel MKL 2019.4.243
\end{itemize}

The median times for the adapted and comparable implementations of the
ATT benchmarks are as follows:

\begin{longtable}[]{@{}
  >{\raggedright\arraybackslash}p{(\columnwidth - 4\tabcolsep) * \real{0.7766}}
  >{\raggedleft\arraybackslash}p{(\columnwidth - 4\tabcolsep) * \real{0.1702}}
  >{\raggedleft\arraybackslash}p{(\columnwidth - 4\tabcolsep) * \real{0.0532}}@{}}
\caption{Matrix calculation}\tabularnewline
\toprule\noalign{}
\begin{minipage}[b]{\linewidth}\raggedright
Operation
\end{minipage} & \begin{minipage}[b]{\linewidth}\raggedleft
Median time (s)
\end{minipage} & \begin{minipage}[b]{\linewidth}\raggedleft
Rank
\end{minipage} \\
\midrule\noalign{}
\endfirsthead
\toprule\noalign{}
\begin{minipage}[b]{\linewidth}\raggedright
Operation
\end{minipage} & \begin{minipage}[b]{\linewidth}\raggedleft
Median time (s)
\end{minipage} & \begin{minipage}[b]{\linewidth}\raggedleft
Rank
\end{minipage} \\
\midrule\noalign{}
\endhead
\bottomrule\noalign{}
\endlastfoot
\(12,000 \times 12,000\) \(\text{matrix}^{1,000}\) - Eigen & \(0.564\) &
1 \\
\(12,000 \times 12,000\) \(\text{matrix}^{1,000}\) - Armadillo &
\(0.763\) & 2 \\
\(12,000 \times 12,000\) \(\text{matrix}^{1,000}\) - R & \(0.988\) &
3 \\
\(14,000 \times 14,000\) cross-product matrix - Armadillo & \(0.47\) &
1 \\
\(14,000 \times 14,000\) cross-product matrix - R & \(0.625\) & 2 \\
\(14,000 \times 14,000\) cross-product matrix - Eigen & \(1.322\) & 3 \\
Creation and modification of a \(12,500 \times 12,500\) matrix -
Armadillo & \(0.321\) & 1 \\
Creation and modification of a \(12,500 \times 12,500\) matrix - Eigen &
\(0.351\) & 2 \\
Creation and modification of a \(12,500 \times 12,500\) matrix - R &
\(0.594\) & 3 \\
Linear regression over a \(15,000 \times 15,000\) matrix - Armadillo &
\(0.616\) & 1 \\
Linear regression over a \(15,000 \times 15,000\) matrix - R & \(8.084\)
& 2 \\
Linear regression over a \(15,000 \times 15,000\) matrix - Eigen &
\(9.604\) & 3 \\
Sorting of 35,000,000 values - Armadillo & \(1.041\) & 1 \\
Sorting of 35,000,000 values - Eigen & \(1.067\) & 2 \\
Sorting of 35,000,000 values - R & \(1.32\) & 3 \\
\end{longtable}

\begin{longtable}[]{@{}
  >{\raggedright\arraybackslash}p{(\columnwidth - 4\tabcolsep) * \real{0.7692}}
  >{\raggedleft\arraybackslash}p{(\columnwidth - 4\tabcolsep) * \real{0.1758}}
  >{\raggedleft\arraybackslash}p{(\columnwidth - 4\tabcolsep) * \real{0.0549}}@{}}
\caption{Matrix functions}\tabularnewline
\toprule\noalign{}
\begin{minipage}[b]{\linewidth}\raggedright
Operation
\end{minipage} & \begin{minipage}[b]{\linewidth}\raggedleft
Median time (s)
\end{minipage} & \begin{minipage}[b]{\linewidth}\raggedleft
Rank
\end{minipage} \\
\midrule\noalign{}
\endfirsthead
\toprule\noalign{}
\begin{minipage}[b]{\linewidth}\raggedright
Operation
\end{minipage} & \begin{minipage}[b]{\linewidth}\raggedleft
Median time (s)
\end{minipage} & \begin{minipage}[b]{\linewidth}\raggedleft
Rank
\end{minipage} \\
\midrule\noalign{}
\endhead
\bottomrule\noalign{}
\endlastfoot
Cholesky decomposition of a \(15,000 \times 15,000\) matrix - Armadillo
& \(0.585\) & 1 \\
Cholesky decomposition of a \(15,000 \times 15,000\) matrix - R &
\(0.819\) & 2 \\
Cholesky decomposition of a \(15,000 \times 15,000\) matrix - Eigen &
\(1.927\) & 3 \\
Determinant of a \(12,500 \times 12,500\) matrix - Armadillo & \(0.387\)
& 1 \\
Determinant of a \(12,500 \times 12,500\) matrix - R & \(0.503\) & 2 \\
Determinant of a \(12,500 \times 12,500\) matrix - Eigen & \(0.635\) &
3 \\
Eigenvalues of a \(3,200 \times 3,200\) matrix - Armadillo & \(0.429\) &
1 \\
Eigenvalues of a \(3,200 \times 3,200\) matrix - R & \(0.464\) & 2 \\
Eigenvalues of a \(3,200 \times 3,200\) matrix - Eigen & \(2.602\) &
3 \\
Fast Fourier Transform over 12,000,000 values - Eigen & \(0.301\) & 1 \\
Fast Fourier Transform over 12,000,000 values - R & \(0.362\) & 2 \\
Fast Fourier Transform over 12,000,000 values - Armadillo & \(0.553\) &
3 \\
Inverse of a \(8,000 \times 8,000\) matrix - Armadillo & \(0.19\) & 1 \\
Inverse of a \(8,000 \times 8,000\) matrix - R & \(0.234\) & 2 \\
Inverse of a \(8,000 \times 8,000\) matrix - Eigen & \(0.448\) & 3 \\
\end{longtable}

\begin{longtable}[]{@{}
  >{\raggedright\arraybackslash}p{(\columnwidth - 4\tabcolsep) * \real{0.7111}}
  >{\raggedleft\arraybackslash}p{(\columnwidth - 4\tabcolsep) * \real{0.2333}}
  >{\raggedleft\arraybackslash}p{(\columnwidth - 4\tabcolsep) * \real{0.0556}}@{}}
\caption{Programmation}\tabularnewline
\toprule\noalign{}
\begin{minipage}[b]{\linewidth}\raggedright
Operation
\end{minipage} & \begin{minipage}[b]{\linewidth}\raggedleft
Median time (s)
\end{minipage} & \begin{minipage}[b]{\linewidth}\raggedleft
Rank
\end{minipage} \\
\midrule\noalign{}
\endfirsthead
\toprule\noalign{}
\begin{minipage}[b]{\linewidth}\raggedright
Operation
\end{minipage} & \begin{minipage}[b]{\linewidth}\raggedleft
Median time (s)
\end{minipage} & \begin{minipage}[b]{\linewidth}\raggedleft
Rank
\end{minipage} \\
\midrule\noalign{}
\endhead
\bottomrule\noalign{}
\endlastfoot
17,500,000 Fibonacci numbers calculation - Armadillo &
\(5.9 \times 10^{-1}\) & 1 \\
17,500,000 Fibonacci numbers calculation - Eigen &
\(5.9 \times 10^{-1}\) & 2 \\
17,500,000 Fibonacci numbers calculation - R & \(7.1 \times 10^{-1}\) &
3 \\
Creation of a \(15,000 \times 15,000\) Hilbert matrix - Eigen &
\(3.1 \times 10^{-2}\) & 1 \\
Creation of a \(15,000 \times 15,000\) Hilbert matrix - Armadillo &
\(3.7 \times 10^{-2}\) & 2 \\
Creation of a \(15,000 \times 15,000\) Hilbert matrix - R &
\(2.4 \times 10^{-1}\) & 3 \\
Creation of a \(2,500 \times 2,500\) Toeplitz matrix - Eigen &
\(3.1 \times 10^{-4}\) & 1 \\
Creation of a \(2,500 \times 2,500\) Toeplitz matrix - Armadillo &
\(5.0 \times 10^{-4}\) & 2 \\
Creation of a \(2,500 \times 2,500\) Toeplitz matrix - R &
\(4.1 \times 10^{-3}\) & 3 \\
Escoufier's method on a \(225 \times 225\) matrix - Eigen &
\(3.4 \times 10^{-2}\) & 1 \\
Escoufier's method on a \(225 \times 225\) matrix - Armadillo &
\(4.8 \times 10^{-2}\) & 2 \\
Escoufier's method on a \(225 \times 225\) matrix - R &
\(3.0 \times 10^{-1}\) & 3 \\
Grand common divisors of 2,000,000 pairs - Armadillo &
\(3.8 \times 10^{-2}\) & 1 \\
Grand common divisors of 2,000,000 pairs - Eigen &
\(4.0 \times 10^{-2}\) & 2 \\
Grand common divisors of 2,000,000 pairs - R & \(2.793\) & 3 \\
\end{longtable}

Repeating the same after multiplicating the number of rows and columns
by five leads to the following results:

\begin{longtable}[]{@{}
  >{\raggedright\arraybackslash}p{(\columnwidth - 4\tabcolsep) * \real{0.7766}}
  >{\raggedleft\arraybackslash}p{(\columnwidth - 4\tabcolsep) * \real{0.1702}}
  >{\raggedleft\arraybackslash}p{(\columnwidth - 4\tabcolsep) * \real{0.0532}}@{}}
\caption{Matrix calculation}\tabularnewline
\toprule\noalign{}
\begin{minipage}[b]{\linewidth}\raggedright
Operation
\end{minipage} & \begin{minipage}[b]{\linewidth}\raggedleft
Median time (s)
\end{minipage} & \begin{minipage}[b]{\linewidth}\raggedleft
Rank
\end{minipage} \\
\midrule\noalign{}
\endfirsthead
\toprule\noalign{}
\begin{minipage}[b]{\linewidth}\raggedright
Operation
\end{minipage} & \begin{minipage}[b]{\linewidth}\raggedleft
Median time (s)
\end{minipage} & \begin{minipage}[b]{\linewidth}\raggedleft
Rank
\end{minipage} \\
\midrule\noalign{}
\endhead
\bottomrule\noalign{}
\endlastfoot
\(12,000 \times 12,000\) \(\text{matrix}^{1,000}\) - Eigen & \(13.974\)
& 1 \\
\(12,000 \times 12,000\) \(\text{matrix}^{1,000}\) - Armadillo &
\(18.988\) & 2 \\
\(12,000 \times 12,000\) \(\text{matrix}^{1,000}\) - R & \(24.48\) &
3 \\
\(14,000 \times 14,000\) cross-product matrix - Armadillo & \(13.78\) &
1 \\
\(14,000 \times 14,000\) cross-product matrix - R & \(17.7\) & 2 \\
\(14,000 \times 14,000\) cross-product matrix - Eigen & \(127.95\) &
3 \\
Creation and modification of a \(12,500 \times 12,500\) matrix -
Armadillo & \(8.037\) & 1 \\
Creation and modification of a \(12,500 \times 12,500\) matrix - Eigen &
\(8.797\) & 2 \\
Creation and modification of a \(12,500 \times 12,500\) matrix - R &
\(13.98\) & 3 \\
Linear regression over a \(15,000 \times 15,000\) matrix - Armadillo &
\(16.634\) & 1 \\
Linear regression over a \(15,000 \times 15,000\) matrix - R & \(1265\)
& 2 \\
Linear regression over a \(15,000 \times 15,000\) matrix - Eigen &
\(1517.8\) & 3 \\
Sorting of 35,000,000 values - Armadillo & \(5.556\) & 1 \\
Sorting of 35,000,000 values - Eigen & \(5.708\) & 2 \\
Sorting of 35,000,000 values - R & \(6.952\) & 3 \\
\end{longtable}

\begin{longtable}[]{@{}
  >{\raggedright\arraybackslash}p{(\columnwidth - 4\tabcolsep) * \real{0.7692}}
  >{\raggedleft\arraybackslash}p{(\columnwidth - 4\tabcolsep) * \real{0.1758}}
  >{\raggedleft\arraybackslash}p{(\columnwidth - 4\tabcolsep) * \real{0.0549}}@{}}
\caption{Matrix functions}\tabularnewline
\toprule\noalign{}
\begin{minipage}[b]{\linewidth}\raggedright
Operation
\end{minipage} & \begin{minipage}[b]{\linewidth}\raggedleft
Median time (s)
\end{minipage} & \begin{minipage}[b]{\linewidth}\raggedleft
Rank
\end{minipage} \\
\midrule\noalign{}
\endfirsthead
\toprule\noalign{}
\begin{minipage}[b]{\linewidth}\raggedright
Operation
\end{minipage} & \begin{minipage}[b]{\linewidth}\raggedleft
Median time (s)
\end{minipage} & \begin{minipage}[b]{\linewidth}\raggedleft
Rank
\end{minipage} \\
\midrule\noalign{}
\endhead
\bottomrule\noalign{}
\endlastfoot
Cholesky decomposition of a \(15,000 \times 15,000\) matrix - Armadillo
& \(16.821\) & 1 \\
Cholesky decomposition of a \(15,000 \times 15,000\) matrix - R &
\(21.027\) & 2 \\
Cholesky decomposition of a \(15,000 \times 15,000\) matrix - Eigen &
\(184.11\) & 3 \\
Determinant of a \(12,500 \times 12,500\) matrix - Armadillo &
\(10.004\) & 1 \\
Determinant of a \(12,500 \times 12,500\) matrix - R & \(12.981\) & 2 \\
Determinant of a \(12,500 \times 12,500\) matrix - Eigen & \(40.249\) &
3 \\
Eigenvalues of a \(3,200 \times 3,200\) matrix - Armadillo & \(9.11\) &
1 \\
Eigenvalues of a \(3,200 \times 3,200\) matrix - R & \(9.445\) & 2 \\
Eigenvalues of a \(3,200 \times 3,200\) matrix - Eigen & \(519.36\) &
3 \\
Fast Fourier Transform over 12,000,000 values - Eigen & \(1.75\) & 1 \\
Fast Fourier Transform over 12,000,000 values - R & \(2.333\) & 2 \\
Fast Fourier Transform over 12,000,000 values - Armadillo & \(3.529\) &
3 \\
Inverse of a \(8,000 \times 8,000\) matrix - Armadillo & \(5.077\) &
1 \\
Inverse of a \(8,000 \times 8,000\) matrix - R & \(6.846\) & 2 \\
Inverse of a \(8,000 \times 8,000\) matrix - Eigen & \(36.693\) & 3 \\
\end{longtable}

\begin{longtable}[]{@{}
  >{\raggedright\arraybackslash}p{(\columnwidth - 4\tabcolsep) * \real{0.7111}}
  >{\raggedleft\arraybackslash}p{(\columnwidth - 4\tabcolsep) * \real{0.2333}}
  >{\raggedleft\arraybackslash}p{(\columnwidth - 4\tabcolsep) * \real{0.0556}}@{}}
\caption{Programmation}\tabularnewline
\toprule\noalign{}
\begin{minipage}[b]{\linewidth}\raggedright
Operation
\end{minipage} & \begin{minipage}[b]{\linewidth}\raggedleft
Median time (s)
\end{minipage} & \begin{minipage}[b]{\linewidth}\raggedleft
Rank
\end{minipage} \\
\midrule\noalign{}
\endfirsthead
\toprule\noalign{}
\begin{minipage}[b]{\linewidth}\raggedright
Operation
\end{minipage} & \begin{minipage}[b]{\linewidth}\raggedleft
Median time (s)
\end{minipage} & \begin{minipage}[b]{\linewidth}\raggedleft
Rank
\end{minipage} \\
\midrule\noalign{}
\endhead
\bottomrule\noalign{}
\endlastfoot
17,500,000 Fibonacci numbers calculation - Armadillo & \(2.893\) & 1 \\
17,500,000 Fibonacci numbers calculation - Eigen & \(2.97\) & 2 \\
17,500,000 Fibonacci numbers calculation - R & \(3.576\) & 3 \\
Creation of a \(15,000 \times 15,000\) Hilbert matrix - Eigen &
\(9.2 \times 10^{-1}\) & 1 \\
Creation of a \(15,000 \times 15,000\) Hilbert matrix - Armadillo &
\(1.062\) & 2 \\
Creation of a \(15,000 \times 15,000\) Hilbert matrix - R & \(4.428\) &
3 \\
Creation of a \(2,500 \times 2,500\) Toeplitz matrix - Eigen &
\(1.8 \times 10^{-2}\) & 1 \\
Creation of a \(2,500 \times 2,500\) Toeplitz matrix - Armadillo &
\(2.9 \times 10^{-2}\) & 2 \\
Creation of a \(2,500 \times 2,500\) Toeplitz matrix - R &
\(1.6 \times 10^{-1}\) & 3 \\
Escoufier's method on a \(225 \times 225\) matrix - Armadillo &
\(25.369\) & 1 \\
Escoufier's method on a \(225 \times 225\) matrix - Eigen & \(38.911\) &
2 \\
Escoufier's method on a \(225 \times 225\) matrix - R & \(403.24\) &
3 \\
Grand common divisors of 2,000,000 pairs - Armadillo &
\(1.9 \times 10^{-1}\) & 1 \\
Grand common divisors of 2,000,000 pairs - Eigen &
\(2.0 \times 10^{-1}\) & 2 \\
Grand common divisors of 2,000,000 pairs - R & \(15.715\) & 3 \\
\end{longtable}

The results are consistent with the local benchmarks, and Armadillo
leads in most of the tests. The benchmarks are also consistent with the
time complexity of the algorithms, meaning that doubling the size of the
matrix does not double the time to run the function unless the function
has a time complexity of \(O(n)\).

\section{Benchmarks with data
transfer}\label{benchmarks-with-data-transfer}

Psarras, Barthels, and Bientinesi (2022) provides different benchmarks
for the Linear Algebra Mapping Problem. We have adapted their benchmarks
to solve linear systems in order to extrapolate their findings to a
larger input data.

We repeated the experiment consisting in solving a linear system of
equations \(AX = B\) with \(A \in \mathbb{R}^{n \times n}\) and
\(B \in \mathbb{R}^{n \times m}\) for \(n = 30,000\) and \(m = 1,000\).
Because only \texttt{cpp11armadillo} has available methods to pass
sparse matrices between R and C++, we created dense matrices to also
include \texttt{cpp11eigen} in the comparison, and this also adds
additional stress to the tests.

A dense matrix with double precision entries
\(A \in \mathbb{R}^{30,000 \times 30,000}\) uses 6.7 GB of RAM and can
be created and operated in the Niagara cluster without problems. The
benchmark repeats the same task in two different ways, in a naive way by
directly computing \(A^{-1}B\) and in a smart way by using
\texttt{solve(A,B)}. The solution always exists because the data is
created by first defining random matrices \(A\) and \(X\) and then
computing \(B = AX\), the benchmarks use \(A\) and \(B\) as inputs to
solve for \(X\). For the cases where \(A^{-1}\) is not directly
obtained, R and Armadillo use the LU decomposition by inspecting the
matrix structure while in Eigen this has to be specified, and this is
the correct approach provided that \(A\) was created with the
\texttt{rnorm()} function in R without guarantee of symmetry nor it
results in an overdetermined system.

The benchmark results are as follows:

\begin{longtable}[]{@{}lrr@{}}
\caption{Solving linear systems}\tabularnewline
\toprule\noalign{}
Operation & Median time (s) & Rank \\
\midrule\noalign{}
\endfirsthead
\toprule\noalign{}
Operation & Median time (s) & Rank \\
\midrule\noalign{}
\endhead
\bottomrule\noalign{}
\endlastfoot
Naive solution - Armadillo & 18.358 & 1 \\
Naive solution - Base R & 62.138 & 2 \\
Naive solution - Eigen & 1899.225 & 3 \\
Smart solution - Base R & 25.494 & 1 \\
Smart solution - Armadillo & 29.394 & 2 \\
Smart solution - Eigen & 467.490 & 3 \\
\end{longtable}

Armadillo leads in performance for the naive solution. This means that,
even after considering the overhead of data transfer, Armadillo excels
at computationally intensive tasks that involve repeated operations as
it is the case of repeating row and column multiplication to obtain
\(X\), and this benchmark does not cover additional data structures in
C++ that do not exist in R and that provide additional flexibility.

R leads in performance for the smart solution not because of the data
transfer overhead, but because R internally uses LAPACK and BLAS
libraries that are highly optimized for linear algebra operations. In
the particular case of the Niagara cluster, R was compiled against the
Intel MKL library, which is highly optimized for Intel processors and
benefits from the specific processor instructions set. Armadillo and
Eigen in this case also use the Intel MKL library, and for a benchmark
with a smaller size input data (e.g. \(100\times 100\))the overhead of
data transfer and using 40 cores would be higher than the speed gains.

\section{Similarities between C++ libraries and R
packages}\label{similarities-between-c-libraries-and-r-packages}

The syntax and speed differences when using Armadillo or Eigen in C++
posit a similar case to the tradeoff between using \texttt{dplyr} and
\texttt{data.table} (Wickham et al. 2019; Barrett et al. 2024).
\texttt{dplyr} is easier to use but \texttt{data.table} is faster.
\texttt{dplyr} was not designed to be fast but \texttt{data.table} was
not designed to be easy to use. For instance, the code to obtain the
grouped means by number of cylinders in the \texttt{mtcars} dataset is:

\begin{Shaded}
\begin{Highlighting}[]
\CommentTok{\# dplyr}
\NormalTok{mtcars }\SpecialCharTok{\%\textgreater{}\%}
  \FunctionTok{group\_by}\NormalTok{(cyl) }\SpecialCharTok{\%\textgreater{}\%}
  \FunctionTok{summarise\_all}\NormalTok{(mean)}

\CommentTok{\# data.table}
\FunctionTok{as.data.table}\NormalTok{(mtcars)[, }\FunctionTok{lapply}\NormalTok{(.SD, mean), by }\OtherTok{=}\NormalTok{ cyl]}
\end{Highlighting}
\end{Shaded}

The local benchmark for the grouped means reveals that \texttt{dplyr}
has a median time of 2.7 ms and \texttt{data.table} has a median time of
600 \(\mu\)s, meaning that \texttt{dplyr} is four times slower than
\texttt{data.table} at this task. The syntax of \texttt{dplyr} is easier
to understand for non-programmers, but \texttt{data.table} can be
equally expressive for users who are familiar with its syntax.

The tests for Armadillo and Eigen reveal that, for repeated and
computationally intensive tasks, rewriting R code in C++ can lead to
significant performance improvements, but it comes at the cost of
learning a new syntax.

As with \texttt{dplyr} and \texttt{data.table}, the choice between
Armadillo and Eigen depends on the user's needs and preferences. For
instance, Armadillo or Eigen can be ideal to work with a
\(1,000,000 \times 1,000,000\) matrix but R can be more suitable for a
\(1,000 \times 1,000\) matrix, and something similar applies to
\texttt{dplyr} that is suitable for a 2-4 GB CSV files or SQL data but
\texttt{data.table} is more suitable for large datasets (e.g., 100 GB
CSV files).

\section{Cases where Armadillo and Eigen stand
out}\label{cases-where-armadillo-and-eigen-stand-out}

Using Armadillo or Eigen can be particularly useful for functions that
involve nested loops and recursion. If we are going to repeatedly use a
function that requires nested loops or multiple linear algebra
operations, it may be worth rewriting it in C++ using Armadillo or Eigen
instead of using base R. In such cases, the time incurred in learning
the syntax and obtaining the correct function is an investment that pays
off in long run time savings.

For instance, Vargas Sepulveda (2020) uses base R and the Matrix package
to calculate the Balassa index and provides international trade data for
226 countries and 785 exported commodities. A matrix of
\(226\times 785\) does not pose a problem for base R, nor it counts as
big data, but it shows large speed gains when using Armadillo or Eigen.

Let \(X \in \mathbb{R}^{C\times P}\) be a matrix with entries
\(x_{c,p}\) that represents the exports of country \(c\) in product
\(p\), from this matrix the Balassa indices matrix is calculated as:

\begin{equation}
\label{eq:balassa1}
B = ([X \oslash (X \vec{1}_{P\times 1})]^t \oslash  [X^t \vec{1}_{C\times 1} \oslash (\vec{1}_{C\times 1}^t X \vec{1}_{P\times 1})])^t,
\end{equation}

where \(\oslash\) denotes element-wise division and \(t\) denotes
transposition.

This is the same as the Balassa index for country \(c\) and product
\(p\): \begin{equation}
\label{eq:balassa2}
B_{cp} = \frac{x_{cp}}{\sum_c x_{cp}} / \frac{\sum_p x_{cp}}{\sum_{c}\sum_{p} x_{cp}}.
\end{equation}

\(B\) is often used to produce a zeroes and ones matrix \(S\) defined
as:

\begin{equation}
\label{eq:balassa3}
s_{c,p} = \begin{cases}1 & \text{ if } b_{cp} > 1\cr 0 & \text{ otherwise} \end{cases},
\end{equation}

where a value of one indicates that country \(c\) has a revealed
comparative advantage in product \(p\) and zero otherwise.

\eqref{eq:balassa3} can be implemented in base R as:

\begin{Shaded}
\begin{Highlighting}[]
\NormalTok{balassa\_r }\OtherTok{\textless{}{-}} \ControlFlowTok{function}\NormalTok{(X) \{}
\NormalTok{  B }\OtherTok{\textless{}{-}} \FunctionTok{t}\NormalTok{(}\FunctionTok{t}\NormalTok{(X }\SpecialCharTok{/} \FunctionTok{rowSums}\NormalTok{(X)) }\SpecialCharTok{/}\NormalTok{ (}\FunctionTok{colSums}\NormalTok{(X) }\SpecialCharTok{/} \FunctionTok{sum}\NormalTok{(X)))}
\NormalTok{  B[B }\SpecialCharTok{\textless{}} \DecValTok{1}\NormalTok{] }\OtherTok{\textless{}{-}} \DecValTok{0}
\NormalTok{  B[B }\SpecialCharTok{\textgreater{}=} \DecValTok{1}\NormalTok{] }\OtherTok{\textless{}{-}} \DecValTok{1}
\NormalTok{  B}
\NormalTok{\}}
\end{Highlighting}
\end{Shaded}

The C++ code using \texttt{cpp11armadillo} is:

\begin{Shaded}
\begin{Highlighting}[]
\PreprocessorTok{\#include }\ImportTok{\textless{}cpp11.hpp\textgreater{}}
\PreprocessorTok{\#include }\ImportTok{\textless{}cpp11armadillo.hpp\textgreater{}}

\KeywordTok{using} \KeywordTok{namespace}\NormalTok{ cpp11}\OperatorTok{;}
\KeywordTok{using} \KeywordTok{namespace}\NormalTok{ arma}\OperatorTok{;}

\OperatorTok{[[}\AttributeTok{cpp11}\OperatorTok{::}\AttributeTok{register}\OperatorTok{]]}\NormalTok{ doubles\_matrix}\OperatorTok{\textless{}\textgreater{}} \VariableTok{balassa\_arma\_}\OperatorTok{(}
  \AttributeTok{const}\NormalTok{ doubles\_matrix}\OperatorTok{\textless{}\textgreater{}\&}\NormalTok{ x}\OperatorTok{)} \OperatorTok{\{}
\NormalTok{  mat X }\OperatorTok{=}\NormalTok{ as\_Mat}\OperatorTok{(}\NormalTok{x}\OperatorTok{);}

\NormalTok{  mat B }\OperatorTok{=}\NormalTok{ X}\OperatorTok{.}\NormalTok{each\_col}\OperatorTok{()} \OperatorTok{/}\NormalTok{ sum}\OperatorTok{(}\NormalTok{X}\OperatorTok{,} \DecValTok{1}\OperatorTok{);}
\NormalTok{  B }\OperatorTok{=}\NormalTok{ B}\OperatorTok{.}\NormalTok{each\_row}\OperatorTok{()} \OperatorTok{/} \OperatorTok{(}\NormalTok{sum}\OperatorTok{(}\NormalTok{X}\OperatorTok{,} \DecValTok{0}\OperatorTok{)} \OperatorTok{/}\NormalTok{ accu}\OperatorTok{(}\NormalTok{X}\OperatorTok{));}
\NormalTok{  B}\OperatorTok{.}\NormalTok{elem}\OperatorTok{(}\NormalTok{find}\OperatorTok{(}\NormalTok{B }\OperatorTok{\textless{}} \DecValTok{1}\OperatorTok{)).}\NormalTok{zeros}\OperatorTok{();}
\NormalTok{  B}\OperatorTok{.}\NormalTok{elem}\OperatorTok{(}\NormalTok{find}\OperatorTok{(}\NormalTok{B }\OperatorTok{\textgreater{}=} \DecValTok{1}\OperatorTok{)).}\NormalTok{ones}\OperatorTok{();}

  \ControlFlowTok{return}\NormalTok{ as\_doubles\_matrix}\OperatorTok{(}\NormalTok{B}\OperatorTok{);}
\OperatorTok{\}}
\end{Highlighting}
\end{Shaded}

The C++ code using \texttt{cpp11eigen} is:

\begin{Shaded}
\begin{Highlighting}[]
\PreprocessorTok{\#include }\ImportTok{\textless{}cpp11.hpp\textgreater{}}
\PreprocessorTok{\#include }\ImportTok{\textless{}cpp11eigen.hpp\textgreater{}}

\KeywordTok{using} \KeywordTok{namespace}\NormalTok{ cpp11}\OperatorTok{;}
\KeywordTok{using} \KeywordTok{namespace}\NormalTok{ Eigen}\OperatorTok{;}

\OperatorTok{[[}\AttributeTok{cpp11}\OperatorTok{::}\AttributeTok{register}\OperatorTok{]]}\NormalTok{ doubles\_matrix}\OperatorTok{\textless{}\textgreater{}} \VariableTok{balassa\_eigen\_}\OperatorTok{(}
  \AttributeTok{const}\NormalTok{ doubles\_matrix}\OperatorTok{\textless{}\textgreater{}\&}\NormalTok{ x}\OperatorTok{)} \OperatorTok{\{}
\NormalTok{  MatrixXd X }\OperatorTok{=}\NormalTok{ as\_Matrix}\OperatorTok{(}\NormalTok{x}\OperatorTok{);}

\NormalTok{  MatrixXd B }\OperatorTok{=}\NormalTok{ X}\OperatorTok{.}\NormalTok{array}\OperatorTok{().}\NormalTok{rowwise}\OperatorTok{()} \OperatorTok{/}\NormalTok{ X}\OperatorTok{.}\NormalTok{rowwise}\OperatorTok{().}\NormalTok{sum}\OperatorTok{().}\NormalTok{array}\OperatorTok{();}
\NormalTok{  B }\OperatorTok{=}\NormalTok{ B}\OperatorTok{.}\NormalTok{array}\OperatorTok{().}\NormalTok{colwise}\OperatorTok{()} \OperatorTok{/} \OperatorTok{(}\NormalTok{X}\OperatorTok{.}\NormalTok{colwise}\OperatorTok{().}\NormalTok{sum}\OperatorTok{().}\NormalTok{array}\OperatorTok{()} \OperatorTok{/}\NormalTok{ X}\OperatorTok{.}\NormalTok{sum}\OperatorTok{());}
\NormalTok{  B }\OperatorTok{=} \OperatorTok{(}\NormalTok{B}\OperatorTok{.}\NormalTok{array}\OperatorTok{()} \OperatorTok{\textless{}} \DecValTok{1}\OperatorTok{).}\NormalTok{select}\OperatorTok{(}\DecValTok{0}\OperatorTok{,}\NormalTok{ B}\OperatorTok{);}
\NormalTok{  B }\OperatorTok{=} \OperatorTok{(}\NormalTok{B}\OperatorTok{.}\NormalTok{array}\OperatorTok{()} \OperatorTok{\textgreater{}=} \DecValTok{1}\OperatorTok{).}\NormalTok{select}\OperatorTok{(}\DecValTok{1}\OperatorTok{,}\NormalTok{ B}\OperatorTok{);}

  \ControlFlowTok{return}\NormalTok{ as\_doubles\_matrix}\OperatorTok{(}\NormalTok{B}\OperatorTok{);}
\OperatorTok{\}}
\end{Highlighting}
\end{Shaded}

If we use UN COMTRADE data for the year 2020 for 234 countries and 5,386
countries in the finest granularity level data from United Nations
(2023), we can observe that Armadillo and Eigen are around two times
faster than base R at obtaining the Balassa matrix, and this includes
the time to move the data between R and C++:

\begin{longtable}[]{@{}lrr@{}}
\caption{Balassa indices}\tabularnewline
\toprule\noalign{}
Operation & Median time (s) & Rank \\
\midrule\noalign{}
\endfirsthead
\toprule\noalign{}
Operation & Median time (s) & Rank \\
\midrule\noalign{}
\endhead
\bottomrule\noalign{}
\endlastfoot
Balassa indices Eigen & 0.013 & 1 \\
Balassa indices Armadillo & 0.014 & 2 \\
Balassa indices R & 0.026 & 3 \\
\end{longtable}

The rest of the methods in Vargas Sepulveda (2020) involve recursion and
eigenvalues computation, and these tasks were already covered in the ATT
benchmark, meaning that the same speed gains can be expected as in the
Balassa matrix.

\section{Considerations}\label{considerations}

Being fair, R was not designed to be fast, and it is not a low-level
language like C++. R is a high-level language that some users consider
easy to learn and use, and it is particularly useful for data
manipulation and visualization. R has a large number of packages that
can be used to conduct a wide range of statistical analyses, and it is
particularly useful for non-programmers.

Armadillo and Eigen are not designed to be easy to use, and they involve
a learning curve to use them effectively. These libraries are
particularly useful for computationally intensive tasks that involve
nested loops and recursion. These languages have speed advantages over R
for two main reasons, that the time it takes for each step in a loop is
shorter in C++ than in R, and that these libraries provide efficient
data structures for vectors and matrices besides providing internal
methods that combine operations to increase speed and reduce memory
usage.

The choice between R and C++ depends on the user's needs and
preferences. In a way, comparing R and C++ is like comparing a Vespa and
a Ducati motorcycle, both are motorcycles but they are designed for
different purposes and excel in different areas. For instance, a Vespa
is ideal for city commuting and a Ducati is ideal for racing, and the
same applies to R and C++.

Using C++ for tasks such as data manipulation and visualization is like
using a Ford F-150 to go to the convenience store, it is feasible but it
is not the most efficient way to do it and the additional fuel
consumption compared to a lighter vehicle ressembles the extra time
required to compile and corrrect the code. In the same way, using R for
computationally intensive tasks is like using a Mini Cooper to load
bricks and construction materials, it is feasible but you cannot expect
the same experience (e.g., speed and comfort) as using a Ford F-150 for
the same task.

Writing proper C++ code requires a good understanding of the syntax and
reading the respective documentation that each library provides. For
instance, without reading the Armadillo documentation, it is very
tempting to transpose a matrix in the following way:

\begin{Shaded}
\begin{Highlighting}[]
\NormalTok{mat }\VariableTok{bad\_transposition\_}\OperatorTok{(}\AttributeTok{const} \DataTypeTok{int}\OperatorTok{\&}\NormalTok{ n}\OperatorTok{)} \OperatorTok{\{}
\NormalTok{  mat a }\OperatorTok{=}\NormalTok{ randn}\OperatorTok{\textless{}}\NormalTok{mat}\OperatorTok{\textgreater{}(}\NormalTok{n}\OperatorTok{,}\NormalTok{ n}\OperatorTok{)} \OperatorTok{/} \DecValTok{10}\OperatorTok{;}
  
\NormalTok{  mat b}\OperatorTok{(}\NormalTok{a}\OperatorTok{.}\NormalTok{n\_cols}\OperatorTok{,}\NormalTok{ a}\OperatorTok{.}\NormalTok{n\_rows}\OperatorTok{);}

  \ControlFlowTok{for} \OperatorTok{(}\DataTypeTok{int}\NormalTok{ i }\OperatorTok{=} \DecValTok{0}\OperatorTok{;}\NormalTok{ i }\OperatorTok{\textless{}}\NormalTok{ a}\OperatorTok{.}\NormalTok{n\_rows}\OperatorTok{;}\NormalTok{ i}\OperatorTok{++)} \OperatorTok{\{}
    \ControlFlowTok{for} \OperatorTok{(}\DataTypeTok{int}\NormalTok{ j }\OperatorTok{=} \DecValTok{0}\OperatorTok{;}\NormalTok{ j }\OperatorTok{\textless{}}\NormalTok{ a}\OperatorTok{.}\NormalTok{n\_cols}\OperatorTok{;}\NormalTok{ j}\OperatorTok{++)} \OperatorTok{\{}
\NormalTok{      b}\OperatorTok{(}\NormalTok{i}\OperatorTok{,}\NormalTok{ j}\OperatorTok{)} \OperatorTok{=}\NormalTok{ a}\OperatorTok{(}\NormalTok{j}\OperatorTok{,}\NormalTok{ i}\OperatorTok{);}
    \OperatorTok{\}}
  \OperatorTok{\}}

  \ControlFlowTok{return}\NormalTok{ b}\OperatorTok{;}
\OperatorTok{\}}
\end{Highlighting}
\end{Shaded}

Or, even worse, trying to make it faster by using OMP, which would be
faster but not efficient in terms of syntax and available alternatives:

\begin{Shaded}
\begin{Highlighting}[]
\NormalTok{mat }\VariableTok{less\_bad\_transposition\_}\OperatorTok{(}\AttributeTok{const} \DataTypeTok{int}\OperatorTok{\&}\NormalTok{ n}\OperatorTok{)} \OperatorTok{\{}
\NormalTok{  mat a }\OperatorTok{=}\NormalTok{ randn}\OperatorTok{\textless{}}\NormalTok{mat}\OperatorTok{\textgreater{}(}\NormalTok{n}\OperatorTok{,}\NormalTok{ n}\OperatorTok{)} \OperatorTok{/} \DecValTok{10}\OperatorTok{;}
  
\NormalTok{  mat b}\OperatorTok{(}\NormalTok{a}\OperatorTok{.}\NormalTok{n\_cols}\OperatorTok{,}\NormalTok{ a}\OperatorTok{.}\NormalTok{n\_rows}\OperatorTok{);}

  \PreprocessorTok{\#ifdef \_OPENMP}
  \PreprocessorTok{\#pragma omp parallel for collapse(2) schedule(static)}
  \PreprocessorTok{\#endif}
  \ControlFlowTok{for} \OperatorTok{(}\DataTypeTok{int}\NormalTok{ i }\OperatorTok{=} \DecValTok{0}\OperatorTok{;}\NormalTok{ i }\OperatorTok{\textless{}}\NormalTok{ a}\OperatorTok{.}\NormalTok{n\_rows}\OperatorTok{;}\NormalTok{ i}\OperatorTok{++)} \OperatorTok{\{}
    \ControlFlowTok{for} \OperatorTok{(}\DataTypeTok{int}\NormalTok{ j }\OperatorTok{=} \DecValTok{0}\OperatorTok{;}\NormalTok{ j }\OperatorTok{\textless{}}\NormalTok{ a}\OperatorTok{.}\NormalTok{n\_cols}\OperatorTok{;}\NormalTok{ j}\OperatorTok{++)} \OperatorTok{\{}
\NormalTok{      b}\OperatorTok{(}\NormalTok{i}\OperatorTok{,}\NormalTok{ j}\OperatorTok{)} \OperatorTok{=}\NormalTok{ a}\OperatorTok{(}\NormalTok{j}\OperatorTok{,}\NormalTok{ i}\OperatorTok{);}
    \OperatorTok{\}}
  \OperatorTok{\}}

  \ControlFlowTok{return}\NormalTok{ b}\OperatorTok{;}
\OperatorTok{\}}
\end{Highlighting}
\end{Shaded}

The correct way to transpose a matrix in Armadillo, such that its
internals apply low-level optimizations, is to use the \texttt{t()}
function, and it also saves time and typing:

\begin{Shaded}
\begin{Highlighting}[]
\NormalTok{mat }\VariableTok{good\_transposition\_}\OperatorTok{(}\AttributeTok{const} \DataTypeTok{int}\OperatorTok{\&}\NormalTok{ n}\OperatorTok{)} \OperatorTok{\{}
\NormalTok{  mat a }\OperatorTok{=}\NormalTok{ randn}\OperatorTok{\textless{}}\NormalTok{mat}\OperatorTok{\textgreater{}(}\NormalTok{n}\OperatorTok{,}\NormalTok{ n}\OperatorTok{)} \OperatorTok{/} \DecValTok{10}\OperatorTok{;}
  \ControlFlowTok{return}\NormalTok{ a}\OperatorTok{.}\NormalTok{t}\OperatorTok{();}
\OperatorTok{\}}
\end{Highlighting}
\end{Shaded}

\section{Conclusion}\label{conclusion}

Armadillo and Eigen can be highly expressive, these are flexible
libraries once the user has learned the syntax, and these languages have
data structures that do not exist in R that help to write efficient
code. Eigen and \texttt{cpp11eigen} do not simplify the process of
writing C++ code for R users but excels at computationally demanding
applications. Armadillo and \texttt{cpp11armadillo}, on the other hand,
provides a balance between speed and ease of use, and it is a good
choice for users who need to write C++ code that is easier to modify and
maintain.

\section{Acknowledgements}\label{acknowledgements}

I appreciate that Paolo Bientinesi from Umea University read a previous
version of this document on arXiv and referred me to Psarras, Barthels,
and Bientinesi (2022).

\section*{References}\label{references}
\addcontentsline{toc}{section}{References}

\phantomsection\label{refs}


\end{document}